\pdfoutput=1 % only if pdf/png/jpg images are used
\documentclass[11pt,a4paper]{article}
\usepackage{jinstpub}

\usepackage{url}
\usepackage[Symbol]{upgreek}

\title{Random on-board pixel sampling (ROPS) X-ray  Camera}

\author[a]{Zhehui Wang,}
\author[a]{O. Iaroshenko,}
\author[a]{S. Li,} %\thanks{Corresponding author.}\\
\author[b]{T. Liu,}
\author[c]{N. Parab,}
\author[d]{W. W. Chen,}
\author[a]{P. Chu,}
\author[a]{G. Kenyon,}
\author[b]{R. Lipton,}
\author[e]{K.-X. Sun}
\affiliation[a]{Los Alamos National Laboratory, Los Alamos, NM 87545, U. S. A.}
\affiliation[b]{Particle Physics Division, Fermilab, Batavia, IL 60510, U. S. A.}
\affiliation[c]{Advanced Photon Source, Argonne National Laboratory, Argonne, IL 60439, U. S. A.}
\affiliation[d]{Purdue University, West Lafayette, IN 47907, U. S. A.}
\affiliation[e]{University of Nevada, Las Vegas, NV 89154, U. S. A.}

\emailAdd{ zwang@lanl.gov}

\abstract{Recent advances in compressed sensing theory and algorithms offer new possibilities for high-speed X-ray camera design. In many CMOS cameras, each pixel has an independent on-board circuit that includes an amplifier, noise rejection, signal shaper, an analog-to-digital converter (ADC), and optional in-pixel storage. When X-ray images are sparse, {\it i.e.}, when one of the following cases is true: (a.) The number of pixels with true X-ray hits is much smaller than the total number of pixels; (b.) The X-ray information is redundant; or (c.) Some prior knowledge about the X-ray images exists, sparse sampling may be allowed. Here we first illustrate the feasibility of random on-board pixel sampling (ROPS) using an existing set of X-ray images, followed by a discussion about signal to noise as a function of pixel size. Next, we describe a possible circuit architecture to achieve random pixel access and in-pixel storage. The combination of a multilayer architecture, sparse on-chip sampling, and computational image techniques, is expected to facilitate the development and applications of high-speed X-ray camera technology.  %{\it \bf First full draft: \underline{Sept. 4, 2016}; Planned submission: Sept. 26, 2016.}
} 


\begin{document}
\maketitle

\section{Introduction}\label{sec:intro}
Quest for ever higher frame rates in imaging motivates new design principles and architectures in camera technology. Examples of high-speed imaging techniques are optical multiplexing and in-pixel storage. The concept of multiplexing may be traced back to Eadweard Muybridge, who demonstrated high-speed imaging by using multiple slow cameras. Each camera took a single image sequentially through delayed timing techniques. Modern optical multiplexing methods, by dividing the incoming light field into multiple views and each view being captured by a separate camera, have achieved frame rates above one billion frames per second and are available commercially. Some limitations of optical multiplexing are: the number of frames is limited by the number of cameras. Light splitting throws away a lot of light. In a 16-way splitting setup, for example, only 1/16 of the light is recorded per frame of image. Furthermore, direct X-ray-splitting has not been used due to the lack of loss-less beam-splitters for X rays. Indirect methods that first convert X rays into visible light are more common in high-speed X-ray imaging. In-pixel storage, also known as in-situ image storage depending on the location of the memories, records multiple frames of image information temporarily before transmitting the stored images off the camera to a permanent depository. In-pixel storage methods have demonstrated frame rates above 10 MHz for hundreds of storage memories per pixel. Higher frame rates above 100 MHz are possible and the feasibilities studies are underway by different groups. When the memories locate inside the pixel, the pixel size limits the number of frames that can be stored. In-pixel storage has been implemented using both charged-coupled devices (CCD) and complementary metal-oxide-semiconductor (CMOS) processes~\cite{Spie:2012}. CCD or CMOS process optimization and pixel architectures will be key to further improvements in sensor response, frame rate, fill factor, pixel resolution, bit depth and other performance parameters.

Here we discuss an approach called random on-board pixel sampling (ROPS) for high-speed X-ray imaging. Recent advances in compressed sensing or compressed sampling (CS) theory and algorithms~\cite{Ela:2010,EK:2012} offer new directions in in-pixel data processing and therefore different approaches to high-speed X-ray camera designs. Under certain conditions, CS theory removes the minimum sampling frequency requirement as stated in the Nyquist-Shannon sampling (NSS) theory. NSS has been the guiding principle for imaging and other data acquisition techniques. However, it gets harder to implement NSS as the imaging frame rate and/or spatial resolution increases. Another indirect consequence of NSS is that the number of X-ray photons increases significantly, which can be difficult to do even with the brightest X-ray sources available. Furthermore, high X-ray doses associated with NSS could also introduce significant changes and even damages to the objects to be imaged. Therefore, removing the minimum sampling frequency requirement in the new CS framework is very attractive. In addition to front-end data reduction, lower X-ray dose requirement, sparse-sampling methods can also be used in conjunction with the optical multiplexing as well as in-pixel storage methods. 

Below, we first give a brief account of the CS principle that ROPS is based upon, since plenty of literature exists even though direct CS applications to high-speed X-ray imaging may be lagging behind. Next, we discuss the data sparsity in X-ray images. Dynamic scenes can have higher signal sparsity than static scenes, a subject that will be explored further elsewhere. We then discuss how to obtain sparse X-ray data through the use of a multilayer detector architecture. Similar to the earlier multilayer concepts using silicon sensor and `4H' cameras concepts using GaAs:Cr sensor~\cite{Wang:2015,Wang:2016}, we explain that a multilayer camera architecture enhances the signal sparsity in each layer by dividing X-ray signals into multiple layers. We emphasize that the multi-layer concept is particularly suited for high energy X-rays with energies above 20 keV since the X-ray absorption length is long. In addition to silicon and GaAs, several other fast X-ray sensor options are compared.  Incoherent sampling, as required by CS, can be achieved in the pixel or chip level. Access to temporary stored data (electric charges) will be the key to front-end data processing and data compression.

%One of the trends in X-ray diffraction and imaging is the increasingly higher data yield, driven in part by the expansion of X rays for material studies and new material discoveries.  Here materials refer to the subjects not only in a material research laboratory, but also in biology, medicine, geology, art restoration and many other areas of experimental science when understanding of atomic compositions, structures and their dynamics are necessary. Introduction of new material fabrication and processing tools such as additive manufacturing, coupled with the predictive power of computational material science through machine learning for example, offer new possibilities of uncovering novel materials and structures that can be tailored to specific applications and environments. Experimental data from X-ray diffraction, imaging, and other X-ray methods are essential to validate new materials, new fabrication processes, as well as computational predictions. 

\subsection{CS principle for X-ray imaging}
Two  necessary and sufficient conditions to implement CS principle are sparse signals and incoherent sampling~\cite{Ela:2010,EK:2012}. Signal sparsity is base-dependent. One well-known example is the Dirac delta function in time, which is sparse in time bases. It is however not sparse in frequency bases since it contains all of the frequency components when decomposed in Fourier transform. In X-ray imaging, the natural bases are the camera pixels. It is possible to achieve sparse signal condition in the pixel bases by using a multilayer three-dimensional (3D) camera structure described previously~\cite{Wang:2015}. Each layer is sufficiently  thin so that the probability of X-ray capture is low in each pixel and therefore most pixels will have no X-ray hit. Sampling technique based on a threshold may be sufficient to record all the X-ray hits. The number of layers however, can reach several 100s or more, depending on the X-ray fluxes. The second scenario of sparse X-ray signals is the X-ray scattering by periodic structures. X-ray signals scattered off one elementary cell is redundant with respect to the corresponding scattering event off another cell. A third scenario of sparse signal is that some prior knowledge about the scene exists. Such prior knowledge could come from previous experiments and existing data. The existing information can be used as a dictionary for sparse sampling of new experiments. Below, we illustrate the third type of signal sparsity by using X-ray images from dynamically compressed glass balls. Incoherent samplings are obtained through random pixel access, and affected by the signal-to-noise ratio.

\subsection{Sparse signals in X-ray phase contrast imaging}
X-ray phase contrast images of compressed glass balls were obtained using polychromatic, high intensity X-rays at beam line 32-ID-B of the Advanced Photon Source (APS). The fundamental energy of the X-ray beam was peaked at 23.6 KeV (11 mm undulator gap). Commercial solid spherical soda lime glass particles (Cospheric LLC, Santa Barbara, CA) were studied. Two diameter ranges were used: 1.0 -- 1.18 mm and 1.7 -- 2.0 mm. The size of the X-ray beam on the sample was 2560 $\times$ 1600 $\mu$m$^2$. A modified Kolsky bar setup was used to dynamically compress the particles at a nominal velocity of 6 m/s.  Deformation and fracture were studied in-situ using the X-ray phase contrast imaging (PCI). A single crystal Lu$_3$Al$_5$O$_{12}$:Ce scintillator (dimensions: 10 mm $\times$ 10 mm $\times$ 100 $\mu$m) was used to convert the propagated X-ray signal to visible light. The visible light images were recorded using a high-speed camera (Shimadzu Hyper Vision HPV-X2). The frame rate was 2 million frames per second with the exposure time of 200 ns per frame. The resolution of the imaging system was estimated to be 6.4 $\mu$m/pixels and the frame size was 102,000 pixels (400 $\times$ 256). Additional details of the experiments and results can be found in~\cite{Chen:2014,Parab:2016}.

\begin{figure}[hbt]
\centering
\includegraphics[width=.85\textwidth]{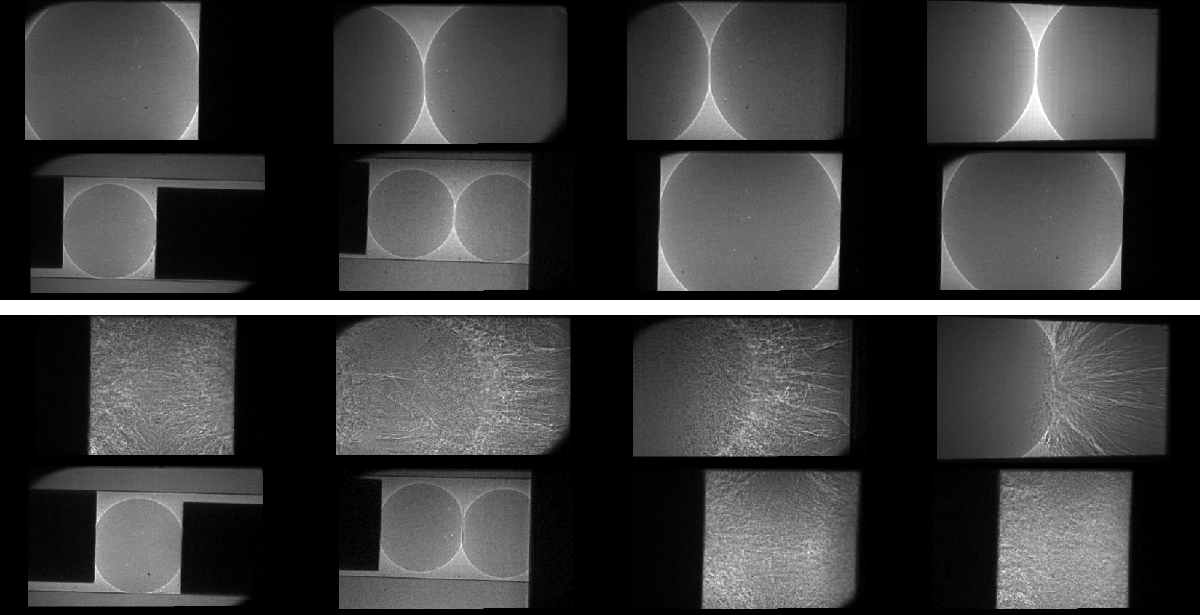}
\caption{Images from eight experiments are used as training data. (Top). Frame 1 out of 256. (Bottom). Frame 176 out of 256.}
\label{fig:train_data1}
\end{figure}

\begin{figure}[hbt]
\centering
\includegraphics[width=.85\textwidth]{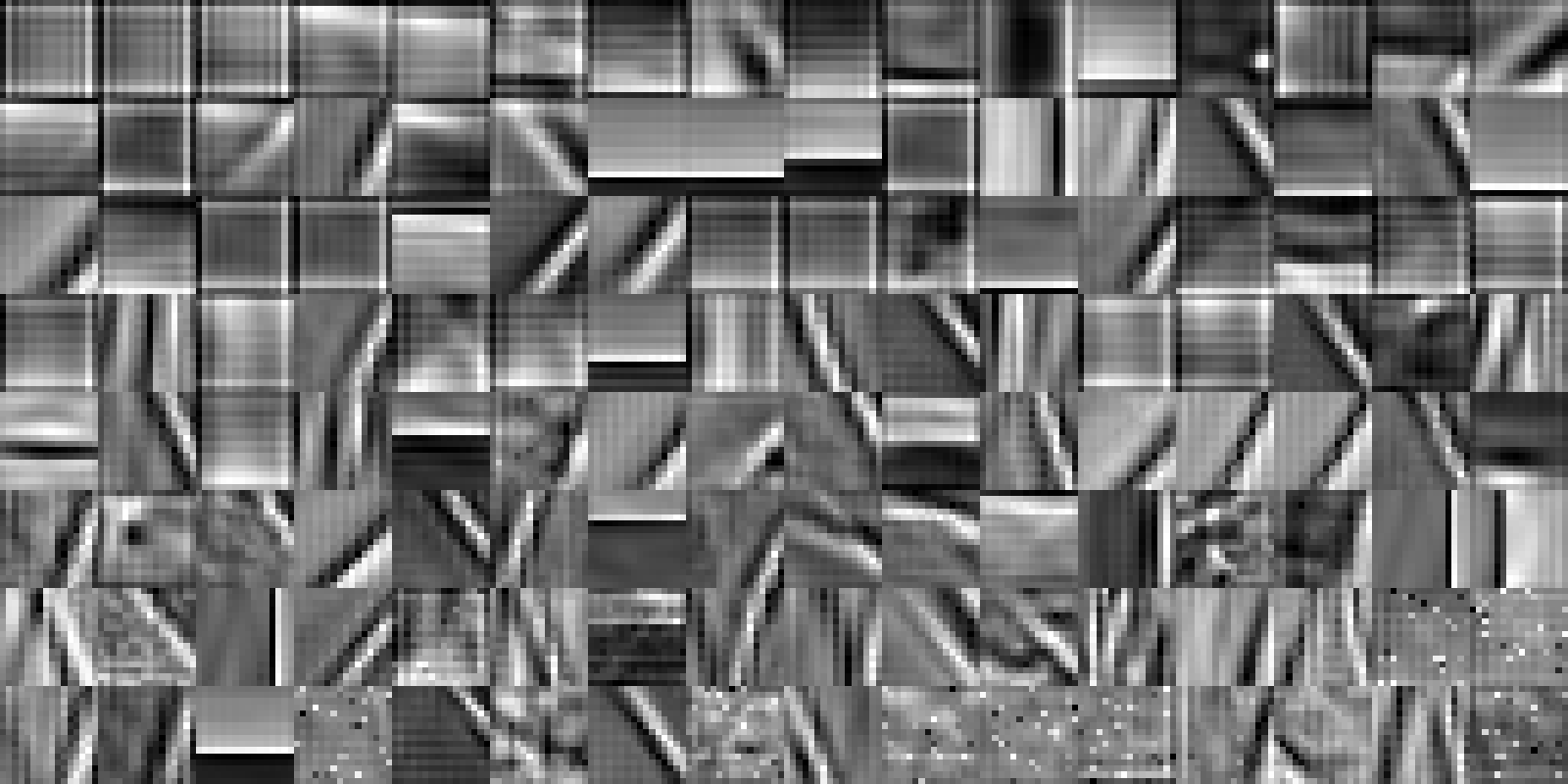}
\caption{128 dictionary elements generated based on the training data, each with a pixel size of 16$\times$16.}
\label{fig:dictionary}
\end{figure}

Here we illustrate the ability of the locally competitive algorithm (LCA) \cite{RozJohBarOls08} to reconstruct an X-ray image from a small sample.  We used the deconvolutional version of LCA \cite{SchPaiLuKen14} implemented in OpenPV \cite{OpenPV}.  First, a sparse code dictionary was learned using data from 8 experiments on crushing balls, each consisting of 256 frames (Fig.~\ref{fig:train_data1}).
In this case we learned 128 dictionary elements of size 16 $\times$16, as shown in Fig.~\ref{fig:dictionary}.
In image reconstruction, Each element can be placed at different positions of the original images with some coefficients. The linear
combinations of all dictionary elements placed at all possible positions allow us to recover the original images. The coefficients in these combinations are a sparse code or sparse representation of an image. Despite the dictionary is over-complete, only a small amount of coefficients are
nonzero. The total number of basis vectors can be calculated as a product
of dictionary size and the number of all possible positions where dictionary elements can be placed within the
image. The LCA achieves sparsity by lateral inhibition among coefficients (also called 'neurons'). When one of the
coefficients becomes nonzero, it suppresses the activity of other neighboring neurons, keeping them in an inactive state or a zero coefficient value.

\begin{figure}[htb]
\centering
\includegraphics[width=.85\textwidth]{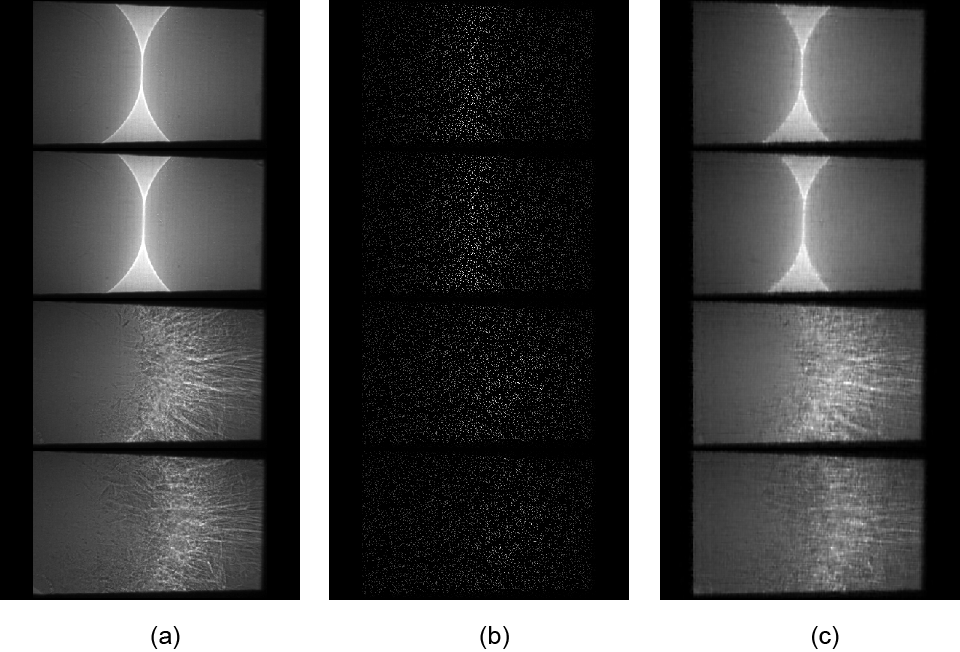}
\caption{(a) Original images at different times; (b) Coarsely sampled images; (c) reconstructed images from the coarsely sampled images using a dictionary generated from other experiments.}
\label{fig:reconstruction}
\end{figure}

After the dictionary has been learned, we use it to reconstruct image frames from two new experiments
given only the masked parts of them. We construct masks to hide 90\% of the pixels chosen randomly
from each original frame. Fig.~\ref{fig:reconstruction} shows the original images, their masked versions, and the reconstructions done
by OpenPV. The sparse representation is lossy, that is, the reconstructed images are not the same as the original ones. But they are clearly identifiable with the original images to the human eyes and therefore preserve some important features from the original images. We also tried 99\% masks that only retain 1\% of the original pixels. The results were too lossy to be useful for reconstruction. A further analysis is to evaluate the use of sparse represented images for the physics of the glass ball compression.

\section{Multilayers \& pixels}
Incoherent sampling may be executed at different stages in the chain of image generation, transmission and recording. In the examples given above, it is executed after the images have been recorded in a permanent depository, a late stage in image registration. This is currently the common approach, which may not be ideal for high speed imaging since as the frame rate increases, the inter-frame time gets shorter and the amount of data can be transmitted off the chip decreases for a fixed data transmission bandwidth. We discuss the frontend approaches equivalent to random pixel sampling as illustrated above, but at the chip level (`on-board') so that significant data reduction may be possible in the first step of image acquisition and recording. Random on-board pixel sampling (ROPS) means that the following two conditions are met: 1.) Every pixel is accessible or addressable for recording; 2.) The probability that a pixel is selected depends on the signal sparsity. In the example given above, the probability is 10\%. Besides the Poisson statistics of detecting X-ray photons, incoherent sampling of signals is affected by noise levels in the pixels. Signals and noise also vary with the X-ray sensor materials, pixel size, electronics noise, storage capacitance, {\it etc.}.

\subsection{Non-uniform-thickness multilayer structure}
\begin{figure}[htb]
\centering
\includegraphics[width=.8\textwidth]{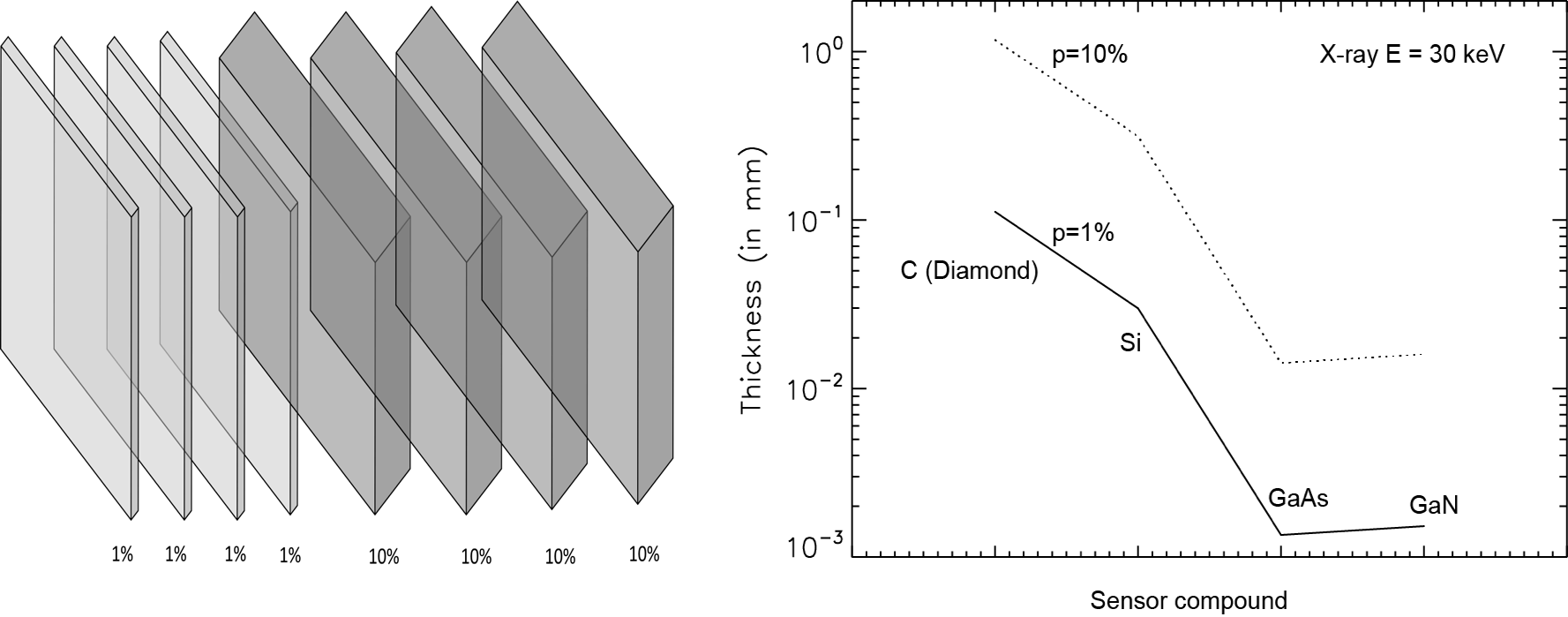}
\caption{(Left). A non-uniform layer thickness design (no to scale) will reduce the individual pixel dynamic range to $\sim$ 10, while allowing the full assembly dynamic range to reach 10$^5$. (Right). Estimated sensor thickness as function of materials for up to 10 photon absorption and 100 photon absorption respectively. Photon energy are assumed to be at 30 keV and incoherent scattering except in diamond may be neglected.}
\label{fig:NUT1}
\end{figure}

Consider X-ray imaging using semiconductor sensors when electron and hole pairs are created primarily by X-ray photoelectric absorptions and contributions from incoherent scatterings are negligible. In other words, Fano-factor-corrected electron-hole pair production is uniform for each photon and the fluctuation is $\sqrt{FN_{eh}}=$32 for a 30 keV photon in silicon ($F=0.12$). Following earlier work~\cite{Wang:2015,Wang:2016}, we continue with the multilayer approaches to X-ray imaging using photon energies at 30 keV and above. Here we would like to introduce a dynamic range constraint per pixel per layer to be around 10. The question is how to design a multilayer detector to allow an {\it equivalent} dynamic range reaching 10$^5$? The term {\it equivalent} refers to a 2D hybrid semiconductor camera with a pixel size of 200 $\mu$m $\times$ 200 $\mu$m and a monolithic sensor thickness of 500 $\mu$m or more. A possible solution is to combine small pixels with thin front-end layers. Compared with a 200 $\mu$m $\times$ 200 $\mu$m pixel, the dynamic range of a 20 $\mu$m $\times$ 20 $\mu$m reduces by 100 to 10$^3$ for the same X-ray flux density. To further reduce the dynamic range per pixel, we may use thin sensors in the frontend (facing the object and incoming beam) layers and thicker layers in the back, as illustrated in Fig.~\ref{fig:NUT1}.

In Fig.~\ref{fig:NUT1}, the dynamic range of 10 or up to 10 X-ray photons per pixel requires that the frontend sensors only absorb 1\% of the maximum flux density of $\sim$ 2.5 ph/$\mu$m$^2$. The number of thin layers is about 10, determined by the statistics as explained below. The thicker backend absorbs about 10\% of the flux per layer. This choice further takes into account the speed requirements that limit the sensor thickness due to the intrinsic material-dependent electron and hole mobilities. If high-speed is not of a concern,  a single monolithic thick layer may be used. However, a multilayer configuration can still help to reduce the dynamic range by spreading out the photons into different layers~\cite{Wang:2016}.

The material-dependent thicknesses shown in Fig.~\ref{fig:NUT1} are the mean absorption thicknesses, derived from the absorption probability $p = 1 - \exp (- \mu_a)$ for $p = 1\%$ and 10\% respectively. $\mu_{a} = n_0 \sigma_a L_a $ is the X-ray attenuation exponent due to absorbing atom number density $n_0$, the photoelectric absorption cross section $\sigma_a$ and absorption layer thickness $L_a$. A fluctuation is expected in the number of detected photons, which satisfies $\Delta N = \sqrt{ \bar{N} ( 1- \bar{N}/N_i)} \sim \sqrt{\bar{N}}$ with $\bar{N}$ being the mean photon counts, which is 10 for 1\% absorption and $100$ for 10\% absorption and $N_i = 10^3$. In actual imaging, $N_i$ is unknown and is to be measured for each pixel. By using a multilayer, the $N_i$ can be estimated more accurately by using the mean of multiple layers or based on the photon count variance. The estimated layer numbers are about 10 for the thin frontend and the thick backend to derive the mean and variance. 
\begin{table}[htb]
\caption{A comparison of X-ray detection using different sensor materials}
\centering
\begin{tabular}{lcccc}
\hline
Properties & Si & GaAs & Diamond & GaN \\
\hline
R$_{CSDA}$(30) ($\mu$m) & 9.91 & 5.73 & 5.62 & 4.51 \\
Fano factor & 0.12 & 0.12& 0.08 & 0.07  \\
breakdown field (V/$\mu$m) & 30 & 10 & $>$10$^2$ & 200   \\
Max. $e^-$ mobility (cm$^2$ V$^{-1}$s$^{-1}$) & 1400 & 8500 &2200&1000  \\
Max. $h^+$ mobility (cm$^2$ V$^{-1}$s$^{-1}$) & 450 & 400 &1800 &350   \\
$e^-$ life time (ms) & $\leq$1 & 10$^{-5}$ & 10$^{-6}$ & 10$^{-6}$\\
$h^+$ life time (ms) & $\leq$1 & 10$^{-6}$ & 10$^{-6}$ & 10$^{-7}$\\
\hline
\end{tabular}
\label{pix:prop}
\end{table}

\subsection{Charge sharing \& pixel noise}
\begin{figure}
\centering
\includegraphics[width=.85\textwidth]{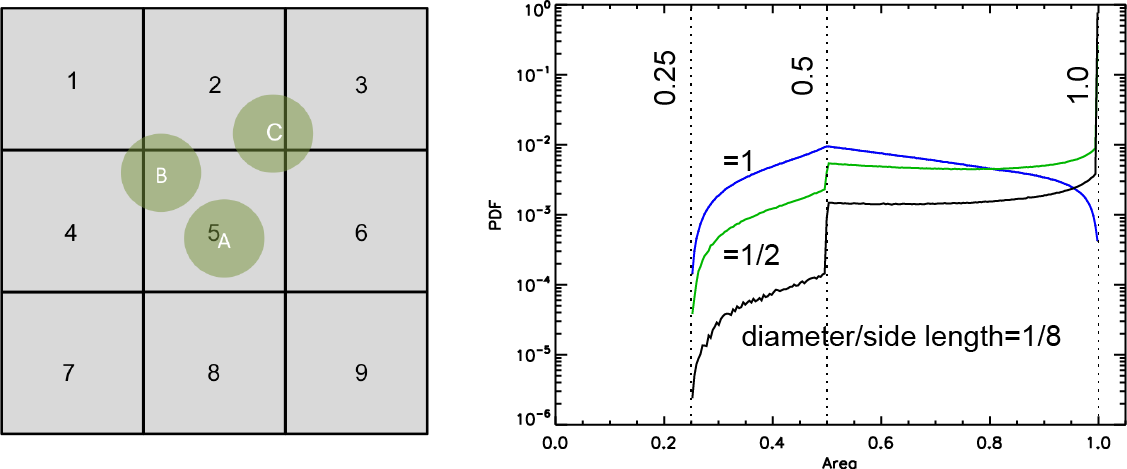}
\caption{(Left). We use colored circles to illustrate charge depositions of monochromatic X-rays (labelled A, B and C) in square pixels (numbered 1 to 9). X-ray-induced charge (B and C) only partially deposits in pixel 5. (Right). Probability density functions (PDFs) of energy deposition when the center of the charge cloud resides in pixel 5. The ratios of charged cloud diameter to the pixel size are 1, 1/2 and 1/8. The fractional charge deposition is proportional to the area of the circle that overlaps with the square pixel.}
\label{fig:ChargeMap1}
\end{figure}

Besides better spatial resolution, it is shown above that using smaller pixels is beneficial to dynamic range reduction. Smaller pixels can also have smaller pixel capacitances (inter-pixel capacitance can now become significant) therefore beneficial to fixed pattern noise (FPN) reduction. In CMOS imagers, FPN, or the variation in pixel outputs in response to the same illumination, originates from device and interconnect mismatches across an image sensor and can not be avoided~\cite{GFH:1998}. The number of in-pixel storage cells also reduces as the pixel size shrinks (3D stacked memories may by-pass this limit). Meanwhile, charge-sharing problem~\cite{PBL:2011} among different pixels become important for small pixel since the photo-electron range can be comparable to the pixel size. Estimates of the 30-keV electron range using continuous slowing-down approximation (CSDA) are given in Table.~\ref{pix:prop}. We adopt the following assumptions for charge sharing study: 1. Ignore electron range straggling; 2. The charge cloud generated by the electron stopping is uniform within the continuous slowing-down approximation (CSDA) radius ($R_{CSDA}$); 3. The photo-electron energy is mono-energetic. In assumption 2, we neglect the charge cloud expansion due to drift. The mean charge cloud spread due to drift is estimated to be $\Delta x = d \sqrt{2\frac{kT}{eV_b}}$, with $b$ is the drift distance. $kT$ is the thermal energy of the electron and $V_b$ the bias voltage. The estimate cloud expansion $\Delta x$ is below a micron. Notice the charge cloud expansion is independent of the mobility of the charge carrier~\cite{Spie:2012}. For a pixel size of $l_p$, the fraction of the electrons that deposit their full energies within the pixel is $(l_p-2R_{CSDA})^2/l_p^2$. If $l_p \leq 2R_{CSDA}$, then essentially no photo-electrons deposit its full energy in a single pixel. For 30 keV X-rays, the condition $l_p \leq 2R_{CSDA}$ is met when $l_p$ is around 20 $\mu$m in silicon. Estimates for other sensor materials can be obtained using the numbers given in Table~\ref{pix:prop}. The energy deposition map per pixel per X-ray as a function of its position in the pixel reduces to a geometric problem, as shown in Fig.~\ref{fig:ChargeMap1}. Due to the symmetries of a square pixel, which is centered at ($0,0$), we only need to consider the quadrant $0 \leq x_0 \leq \frac{l_p}{2}$,  $0 \leq y_0 \leq \frac{l_p}{2}$. 

One of the concerns with charge sharing is double counting of the photons. Another concern is the reduction of signal to noise ratio due to the sharing of X-ray-induced charge among neighboring pixels. In additional to FPN, operation-related noise includes kTC thermal noise, pre-amplifier noise, correlated double sampling noise (CDS), readout noise, dark current, charge-transfer or reset noise (at least one-charge transfer is needed in a CMOS pixel~\cite{Fossum:2014}). Electron velocity and number fluctuations are the origins of the different types of noise~\cite{Spie:2012}.  We estimate that the noise amounts to be 10$^2$ to 10$^3$ elementary charge ($e_0$), smaller noise corresponds to smaller pixel size around 20 $\mu$m~\cite{Lutz:1999}. When the charge is shared among 4 neighboring pixels, the signal is about 2000 $\pm$ 16 $e_0$. A pixel dimension below 100 $\mu$m is necessary to reach a signal to noise ratio of 3 or higher. 

\section{Chip-level incoherent sampling}
The pixel-level circuit functions are summarized here. For ROPS: reset, random pixel access/enabling and selection. To maximize signal-to-noise: a threshold, CDS, amplification. To maximize frame-rate: in-pixel memory elements and buffers to compare the signal frame-to-frame and logics to send out the signal difference could be implemented~\cite{PVG:2013}. Additional common functions include time stamp, address decoding, and readout. It is preferable to have analog-to-digital conversion outside the chip as before~\cite{Wang:2016}. 

\begin{figure}[htb]
\centering
\includegraphics[width=.9\textwidth]{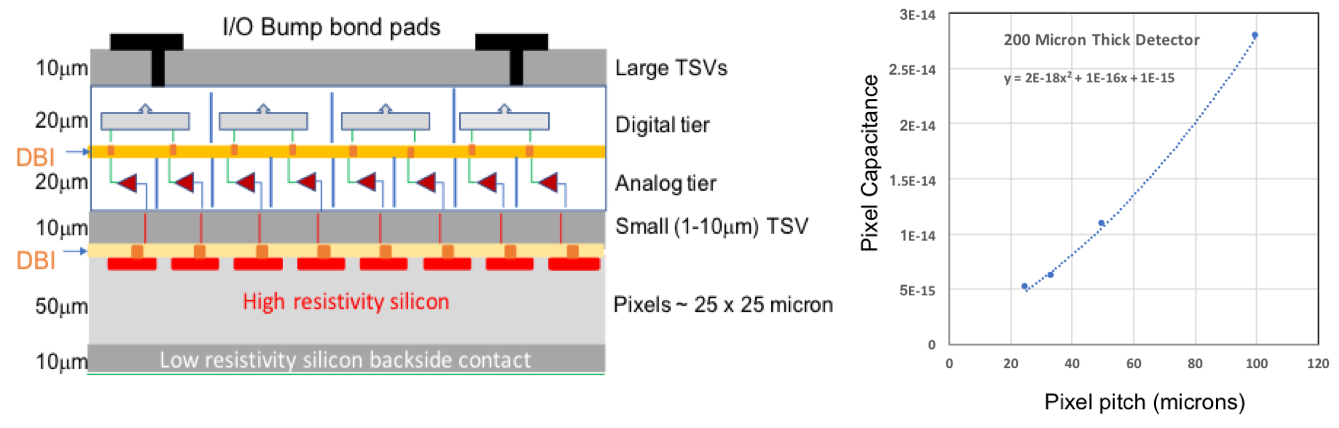}
\caption{(Left) Conceptual pixel design for a three-tier x-ray imaging camera. (Right) Capacitance estimate as a function of pixel size.}
\label{fig:LiuComb}
\end{figure}

A conceptual design of 3D tiered X-ray imaging pixel detector is shown in Fig.~\ref{fig:LiuComb}. The proposal is based on recent work by Fermilab and collaborators~\cite{Liu:2015,Deptuch:2016}. The proposed sensor tier contains sensor with thickness of about 50 $\mu$m (based on demonstrated technology), each pixel size can be as small as 25$\mu$m $\times$ 25$\mu$m. The analog tier contains amplifier and discriminator for each pixel. The digital tier, directly above the analog section, consists of larger cells (100 $\mu$m $\times$ 100 $\mu$m or larger) with each cell capable of processing a field of multiple analog pixels. 

The larger size digital cell allows not only larger local memory storage but also ROPS and other pixel function listed above to be implemented. For example, the top digital tier can correct for charge sharing among pixel and average local timing information to provide improved hit time resolution. The top tier can also provide time stamp and pulse height buffering for high rate single x-ray hit readout. This intrinsically flexible 3D architecture could also open the possibility of implementing motion detection capability based on Associative Memory or CAM~\cite{Liu:2015}, allowing the hit pattern of next frame to be compared with that of the previous frame on the fly in real time. 3DIC technology provides interconnect dimensions small enough to enable pixel pitch below 20 microns.  Fig.~\ref{fig:LiuComb} gives estimates of capacitance as a function of pixel pitch, ignoring interconnect effects. However, the interconnect capacitance is a large contribution to the total load in typical bump bonded pixels and can be dominant for small pixels in conventional technology.  Interconnect load is substantially reduced due to the small TSV dimensions in 3D bonded devices, leading to potential time resolutions of 10's of picoseconds. 

{\it Acknowledgments}
This research used resources of the Advanced Photon Source(APS), a U.S. Department of Energy (DOE) Office of Science User Facility operated for the DOE Office of Science by Argonne National Laboratory under Contract No. DE-AC02-06CH11357. Authors want to thank Kamel Fezzaa, Tao Sun and Alex Deriy at beam line 32-ID, APS for their help in conducting the experiments. The glass particle fracture experiments were funded by ONR Grant N00014-14-1-0628 (Program manager: Dr. David Shifler). Z.W. would like to thank Drs. Cris W. Barnes and Rich Sheffield for stimulating discussions and encouragement to carry out the work.


\begin{thebibliography}{9}
\bibitem{Spie:2012} H. Spieler,
\emph{Electronics and data acquisition.}
{\emph{Nucl. Instrum. Meth. A.} {\bf 666} (2012) 197}.

%\bibitem{Heij:2007} E. H. M. Heijne,
%\emph{Microelectronics technologies for new detectors in medical imaging}
%{\emph{Nucl. Instrum. Meth. A.} {\bf 571} (2007) 7}.

\bibitem{Ela:2010} M. Elad,
\emph{Sparse and Redundant Representations}.
{Springer, 2010}

\bibitem{EK:2012} Y. C. Eldar and G. Kutyniok (eds.)
\emph{Compressed Sensing, theory and applications,}
{(Cambridge, 2012)}

\bibitem{Wang:2015}
Z. Wang, 
\emph{On the Single-Photon-Counting (SPC) modes of imaging using an XFEL source}, { JINST {\bf 10} (2015) C12013}.

\bibitem{Wang:2016}
A. Dragone, C. Kenney, A. Lozinskaya, O. Tolbanov, A. Tyazhev, A. Zarubin and Z. Wang 
\emph{Feasibility study of a `4H' X-ray camera based on GaAs:Cr sensor}, { JINST {\bf 11} (2016) C11042}.

\bibitem{Chen:2014} Chen, W. W., Hudspeth, M. C., Claus, B., Parab, N. D., Black, J. T., Fezza, K. \& Luo, S. N., 
\emph{In situ damage assessment using synchrotron X-rays in materials loaded by a Hopkinson bar.}
{{Phil. Trans. Roy. Soc. A}, 372(2015), 20130191}.

\bibitem{Parab:2016} Parab, N. D., Guo, Z., Hudspeth, M., Claus, B., Lim, et al.,
 \emph{In situ observation of fracture processes in high-strength concretes and limestone using high-speed X-ray phase-contrast imaging.} 
 {Phil. Trans. Roy. Soc. London A. {\bf 375}(2085) (2016).}
 
 \bibitem{RozJohBarOls08} Rozell, C. J.,  Johnson, D. H, Baraniuk, R. G and Olshausen, B. A.
 \emph{Sparse coding via thresholding and local competition in neural circuits, }
 {\emph{Neural computation} {\bf 20}(10), 2526, (MIT Press, 2008).}
 
 \bibitem{SchPaiLuKen14} Schultz, P. F and Paiton, D. M and Lu, W. and Kenyon, G. T.,
 \emph{Replicating kernels with a short stride allows sparse reconstructions with fewer independent kernels.}
 {{arXiv:1406.4205}, (2014).}
 
 \bibitem{OpenPV} \url{http://petavision.github.io/}.
 
\bibitem{GFH:1998} A. E. Gamal, B. Fowler, H. Min, X. Liu,
\emph{Modeling and Estimation of FPN components in CMOS image sensors,}
{\emph{SPIE} 3301 (1998) 168.}

\bibitem{PBL:2011} D. Pennicard, R. Ballabriga, X. LIopart, M. Campbell and H. Graafsma
\emph{Simulations of charge summing and threshold dispersion effects in Medipix3}
{\emph{Nucl. Instrum. Meth. A.} {\bf 636} (2011) 74}.

\bibitem{Fossum:2014} E. R. Fossum and D. B. Hondongwa,
\emph{A Review of the Pinned Photodiode for CCD and CMOS Image Sensors,}
{\emph{IEEE J. Electr. Dev. Soc.} {\bf 2}(3) (2014) 33.}

\bibitem{Lutz:1999} G. Lutz,
\emph{Semiconductor Radiation Detectors: device physics,}
{(Springer, 1999).}

\bibitem{PVG:2013} T. Puthussery, S. Venkataramani, {\it et al.},%J. Gayet-Primo, R. G. Smith and W. R. Taylor, 
\emph{NaV1.1 channels in axon initial segments of bipolar cells augment input to magnocellular visual pathways in the primate retina}, 
{\emph{J. Neurosci} {\bf 33} (2013) 16045}.

\bibitem{Liu:2015} T. Liu, G. Deptuch, J. Hoff, S. Jindariani, {\it et al.},
\emph{Design and Testing of the first 2D prototype VIPRAM}
{JINST 10(02), C02029 (2015)}

\bibitem{Deptuch:2016} G.~W.~Deptuch {\it et al.},
\emph{Fully 3D-Integrated Pixel Detectors for X-Rays}
{IEEE Trans.\ Electron.\ Dev.\  {\bf 63}, 205 (2016).}



%
%\bibitem{BCH:2007} R. Ballabriga, M. Campbell, H. Heijne, X. LIopart, and L. Tlustos
%\emph{The Medipix3 prototype,} % a pixel readout chip working in single photon counting mode with improved spectrometric performance}
%{\emph{IEEE. Trans. Nucl. Sci. } {\bf 54} (2007) 1824}.

%\bibitem{Gruner:2012}S. M. Gruner,
%\emph{X-ray imaging detectors}, 
%{\emph{Physics Today} {\bf 65}(12) (2012) 29}.
%
%\bibitem{HG:2014}T. Hatsui and H. Graafsma,
%\emph{X-ray imaging detectors for synchrotron and XFEL sources},
%{IUCrJ {\bf 2} (2015) 371}.
%
%\bibitem{CSPAD:2012} P. Hart, et al.,
%\emph{Proc. SPIE} {\bf 8504} (2012) 85040.
%
%\bibitem{AGIPD:2014} A. Allahgholi et al., 
%\emph{AGIPD, a high dynamic range fast detector for the European XFEL}, {JINST {\bf 10} (2014) C01023}.
%
%\bibitem{Lucas:2009} L. J. Koerner, M. W. Tate and S. M. Gruner,
%{\emph{IEEE Trans. Nucl. Sci.} {\bf 56} (2009) 2835}.
%
%\bibitem{LPD:2013} A. Koch, M. Hart, T. Nicholls, C. Angelsen, J. Coughlan, M. French, S. Hauf, M. Kuster, J. Sztuk-Dambietz, M. Turcato,
%G. A. Carini, M. Chollet, S. C. Herrmann, H. T. Lemke, S. Nelson, S. Song, M. Weaver, D. Zhu, A. Meents and P. Fischer,
%\emph{Performance of an LPD prototype detector at MHz frame rates underSynchrotron and FEL radiation}, {jinst {\bf 8} (2013) C11001}.
%
%\bibitem{JUNFRAU} J. H. Smith, A. Mozzanica and B. Schmitt, 
%\emph{JUNGFRAU A dynamic gain switching detector for SwissFEL.}
%{\emph{PSI Technical design report} (2015)}.
%
%\bibitem{Veale:2014A}M. C. Veale,  S.J. Bell, D.D. Duarte, M.J. French, A. Schneider, P. Seller, M.D. Wilson, A.D. Lozinskaya, V.A. Novikov, O.P. Tolbanov, A. Tyazhev and A.N. Zarubin,
%\emph{Chromium compensated gallium arsenide detectors for X-ray and $\gamma$-ray spectroscopic imaging}, 
%{\emph{Nucl. Instrum. Meth. A.} {\bf 752}, 6 (2014)}.
%
%\bibitem{Thornber:1974} K. K. Thornber
%\emph{Theory of Noise in Charge-Transfer Devices,}
%{\emph{Bell Syst. Technol. J.} {\bf 53}(7), 1211 (1974).}
%
%\bibitem{JSB:2016} J. H. Jungmann-Smith, A. Bergamaschi, {\it et al.},
%\emph{Towards hybrid pixel detectors for energy-dispersive or soft X-ray photon science,}
%{\emph{J. Synch. Rad.} {\bf 23} (2016) 385.}
%
%\bibitem{Hamann:2014} E. Hamann, T. Koenig, M. Zuber, A. Cecilia, A. Tyazhev, O. Tolbanov, S. Procz, A. Fauler, T. Baumbach, and M. Fiederle,
%\emph{Performance of a Medipix3RX Spectroscopic Pixel Detector With a High Resistivity Gallium Arsenide Sensor}, 
%{\emph{IEEE Trans. Med. Imag.} {\bf 34}, 707 (2015)}.
%
%\bibitem{BTN:2014} D. Budnitsky, A.Tyazhev, V. Novikov, A. Zarubin, O. Tolbanov, M. Skakunov, E. Hamann, A. Fauler, M. Fiederle, S. Procz, 
%\emph{Chromium-compensated GaAs detector material and sensors}, \jinst{9}{2014}{C07011}.
%
%\bibitem{Ayzenshtat:2002} G. I. Ayzenshtat, D. L. Budnitsky, O. B. Koretskay, V. A.  Novikov, L. S. Okaevich, A. I. Potapov, O. P. Tolbanov, A. V. Tyazhev, A. P. Vorobiev.
%\emph{GaAs resistor structures for X-ray imaging detectors}, 
%{\emph{Nucl. Instrum. Meth. A.} {\bf 487}, 96 (2002)}.
%
%\bibitem{Tyazhev:2003} A. V. Tyazhev, D. L. Budnitsky, O. B. Koretskay, V. A.  Novikov, L. S. Okaevich, A. I. Potapov, O. P. Tolbanov, A. P. Vorobiev.
%\emph{GaAs radiation imaging detectors with an active layer thickness up to 1mm}, 
%{\emph{Nucl. Instrum. Meth. A.} {\bf 509}, 34 (2003)}.
%
%\bibitem{Tyazhev:2005} A. V. Tyazhev, V. A.  Novikov, O. B. Koretskaya, D. Y. Mokeev, O. P. Tolbanov, S. A. Ryabkov, G. I. Ayzenshtat
%\emph{GaAs radiation imaging detectors for nondestructive testing, medical, and biological applications}
%{\emph{Proc. SPIE.} {\bf 5922} (2005) 59220Q}.
%
%\bibitem{Tyazhev:2014} A. V. Tyazhev, V. Novikov, O. Tolbanov, A. Zarubin, M. Fiederle, E. Hamann
%\emph{Investigation of the current-voltage characteristics, the electric field distribution and the charge collection efficiency in x-ray sensors based on chromium compensated gallium arsenide}
%{\emph{Proc. SPIE.} {\bf 9213} (2014) 59220Q}.
%
%\bibitem{ATV:2001} G. I. Ayzenshtat, O. P. Tolbanov, A. P. Vorobiev.
%\emph{Modeling of processes of charge division and collection in GaAs detectors taking into account trapping effects}, 
%{\emph{Nucl. Instrum. Meth. A.} {\bf 466}, 1 (2001)}.
%
%\bibitem{Becker:2010} J. Becker, D. Eckstein, R. Klanner and G. Steinbr\"uck
%\emph{Impact of plasma effects on the performance of silicon sensors at an X-ray FEL}, 
%{\emph{Nucl. Instrum. Meth. A.} {\bf 615}, 230 (2010)}.
%
%\bibitem{Drag:2014} A. Dragone, P. Caragiulo, B. Markovic, {\it et al}., 
%\emph{ePix: a class of architectures for second generation LCLS cameras},
%{\emph{Journal of Physics: Conference Series} {\bf 493}, 012012 (2014)}.
%
%\bibitem{bib3}
%A.I. Harris,
%\emph{Spectroscopy with multichannel correlation radiometers},
%\href{http://dx.doi.org/10.1063/1.1898643}
%{\emph{Rev.\ Sci.\ Instrum.} {\bf 76} (2005) 054503}
%[\astroph{0504449}].
%
%\bibitem{bib4}
%G.F. Knoll, \emph{Radiation detection and measurements}, John Wiley
%    and Sons, Inc., New York 2000.
%
%\bibitem{bib5}
%V. Dangendorf, \emph{Time-resolved fast-neutron imaging with a
%pulse-counting image intensifier}, in proceedings of
%\emph{International workshop on fast neutron detectors and
%applications}, April, 3--6, 2006 University of Cape Town, South Africa
%\pos{PoS(FNDA2006)008}.
%
\end{thebibliography}
\end{document}